\documentclass[namedreferences]{SolarPhysics} 

\usepackage{lscape}
\usepackage{url}

\usepackage{epsfig}
\usepackage[optionalrh]{spr-sola-addons}
\usepackage{longtable}

\newcommand{\degr}{$^{\circ}$}

\begin{document}
\begin{article}
\begin{opening}

\title{Solar Flares and Coronal Mass Ejections: A Statistically Determined Flare Flux-CME Mass Correlation}

\author{A.~N.~Aarnio$^{1}$\footnote{Current address: Department of Astronomy, University of Michigan, 830 Dennison Building, 500 Church Street, Ann Arbor, MI 48109. Email: \url{aarnio@umich.edu}}, K.~G.~Stassun$^{1,2}$, W.~J.~Hughes$^{3}$, S.~L.~McGregor$^{3}$}

\runningauthor{Aarnio et al.}
\runningtitle{Flare-CME Flux-Mass Correlation}

\institute{$^{1}$ Department of Physics and Astronomy, Vanderbilt University, VU Station B 1807, Nashville, TN 37235, USA 
                  email: \url{alicia.n.aarnio@vanderbilt.edu}\\
           $^{2}$ Department of Physics, Fisk University, 1000 17$^{th}$ Avenue N., Nashville, TN, 37208 \\
           $^{3}$ Department of Astronomy and Center for Integrated Space Weather Modeling, Boston University, Boston, MA, USA\\
          }
 
\begin{abstract}

In an effort to examine the relationship between flare flux and corresponding CME 
mass, we temporally and spatially correlate all X-ray flares and CMEs in the LASCO 
and GOES archives from 1996 to 2006. We cross-reference 6 733 CMEs having 
well-measured masses against 12 050 
X-ray flares having position information as determined from their optical 
counterparts.  For a given flare, we search in time for CMEs which occur 10-80 
minutes afterward, and we further require the flare and CME to occur within 
$\pm$45\degr\ in position angle on the solar disk. There are 826 CME/flare 
pairs which fit these criteria.  Comparing the flare fluxes with CME masses of 
these paired events, we find CME mass increases with flare flux, following an 
approximately log-linear, broken relationship: in the limit of lower flare fluxes, 
log(CME mass) $\propto$ 0.68$\times$log(flare flux), and in the limit of higher flare 
fluxes, log(CME mass) $\propto$ 0.33$\times$log(flare flux).  
We show that this broken power-law, and in particular the flatter slope at 
higher flare fluxes, may be due to an observational bias against CMEs associated 
with the most energetic flares: halo CMEs. Correcting for this bias yields a 
single power-law relationship of the form log(CME mass) $\propto$ 0.70$\times$ 
log(flare flux).
This function describes the relationship between CME mass and flare flux over 
at least 3 dex in flare flux, from $\approx$10$^{-7} - $10$^{-4}$ W m$^{-2}$.

\end{abstract}
\end{opening}

\section{Introduction\label{intro}}

The physical connection between solar flares and coronal mass ejections has long been a topic of 
debate and ongoing research in solar physics.  CMEs have been observed to occur in conjunction 
with flares and eruptive prominences \cite{Munro:1979,Webb:1987} and with helmet streamer disruptions 
\cite{Dryer:1996}.  While these phenomena are not causally related, it can be definitively said that 
both X-ray flares and CMEs involve reconfiguration of complex magnetic topologies within the corona 
via magnetic reconnection. Indeed, \citeauthor{Svestka:2001} (\citeyear{Svestka:2001}) 
asserted that CMEs all share the same cause---magnetic field lines opening---and that the factor governing 
the properties of the resulting CME is the magnetic field strength in the region whence the CME 
originates.
\inlinecite{Nindos:2004} proposed that helicity may be the link between flares and CMEs: 
CMEs have been observed to facilitate helicity loss, ``carrying'' high helicity magnetic flux from 
the Sun \cite{Chen:1997,Wood:1999,Vourlidas:2000}; \citeauthor{LaBonte:2007} (\citeyear{LaBonte:2007}) 
observed that active regions producing X-class flares generate enough helicity to match that lost via 
CME within hours to days after the flare.

Efforts have been made to identify a physical link between the two phenomena by correlating properties 
of cospatial, contemporaneous flares and CMEs.  Although flares themselves are 
not the cause of CMEs, their properties could serve as prediction tools for imminent CMEs and particle 
events, which are of much concern in the field of space weather. 

Statistical relationships between solar flares 
and CMEs are of interest to astronomers as large scale time-series X-ray observations have found 
solar-like X-ray activity on T Tauri Stars (hereafter referred to as TTS; solar-mass, pre-main sequence 
stars).  Indeed, \inlinecite{Haisch:1995} noted that stellar X-ray flare light curves behave similarly to solar 
X-ray flares, that is, the structure of the light curve (impulsive initial rise followed by exponential 
decay) is effectively identical, though the TTS flares are often several orders of magnitude more 
energetic. Based on these observations, theory and modeling 
of stellar flares has been guided by the premise that the physics 
behind both solar and stellar X-ray flares is the same (\textit{e.g.}, \opencite{Reale:1997}). \citeauthor{Peres:2001} 
(\citeyear{Peres:2001}) demonstrated that stellar coronal X-ray emission could be reproduced by ``scaling up''
solar observations to include greater fractional coverage of the surface by active regions and solar-like 
coronal structures. Further underscoring the similarities in solar/stellar coronae, features similar to 
helmet streamers and 
slingshot prominences have indeed been found on TTS (\textit{e.g.}, \citeauthor{Massi:2008}, \citeyear{Massi:2008}; 
\citeauthor{Skelly:2008}, \citeyear{Skelly:2008}). 

In many applications of TTS X-ray studies, the observable phenomenon is 
the X-ray flare, but the desired quantity is the associated mass loss.  While solar CMEs do not shed large 
quantities of mass, young Suns ($\approx$1 Myr) exhibit $\approx$3 orders of magnitude more energetic flares 
at a higher frequency, and the problem of how substantial angular momentum vis-\`{a}-vis mass is shed in 
these stars remains unresolved. Additionally, CME-like events on young stars could aid in understanding 
circumstellar disk evolution and planet formation: for example, could CMEs on young stars be a 
mechanism for ``flash-heated'' chondrule formation \cite{Miura:2007}?

To be clear, the motivation for this work is not to suggest a causal solar flare-CME relationship. 
We seek to quantify a relationship between solar flare fluxes and CME masses under the general premise 
that in some cases, flares and CMEs arise from common regions of complex magnetic topology and high 
field strength and thus may have correlated properties.  Specifically, we aim to calibrate a relationship 
between solar flare flux and CME mass for the purpose of application to other stars possessing 
solar-like coronae.

\section{Archival Data\label{data}}

\begin{figure}
\centerline{\includegraphics[width=5.5in]{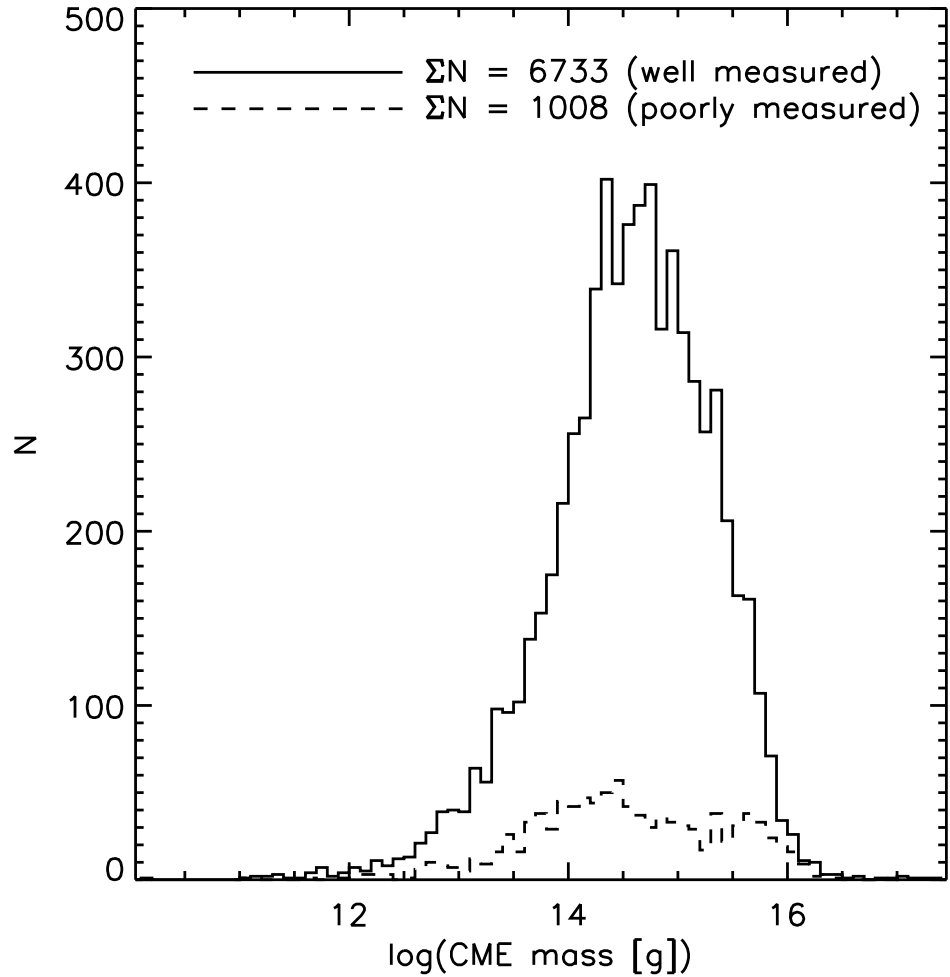}}
\caption{Summary of the LASCO CME database. From 1996 to 2006, there were 7 741
CME mass measurements; 6 733 were well-constrained, and 1 008 were poorly constrained.
}
\label{F-masshist}
\end{figure}

We obtained archival measurements of flares and CMEs for the period of January 1996 to December 2006.
The LASCO CME database \cite{Gopalswamy:2009}\footnote{This CME catalog is generated and 
maintained at the CDAW Data Center by NASA and The Catholic University of America in cooperation 
with the Naval Research Laboratory. SOHO is a project of international cooperation between ESA 
and NASA.} catalogs observations of the Large Angle and Spectrometric Coronagraph dating back to 
January, 1996.  LASCO observations were complete for $\approx$83\% of the 11 years analyzed in this 
work, with 13 862 manually identified CMEs and their measured parameters from this period. Of 
particular interest in this work, 6 733 CMEs with well-measured masses---\textit{i.e.} not halo CMEs or CMEs 
which faded before entering the LASCO field of view---are reported (see Figure \ref{F-masshist}), 
along with their linear speeds, accelerations, angular widths and central position angles (CPAs). 
CME masses are measured by converting brightness per pixel into mass per pixel; this calculation 
assumes the observed brightness is light Thomson scattered off of electrons in a mixture of 90\% 
ionized Hydrogen and 10\% Helium. All pixels within 
a manually defined CME shape are then integrated to obtain a total CME mass \cite{Vourlidas:2000}.
The CME start time reported is when the CME is first detected in the C2 telescope field of view; C2 images 
the region from 2.0 to 6.0 $R_{\odot}$. 

\begin{figure}
\centerline{\includegraphics[width=5.5in]{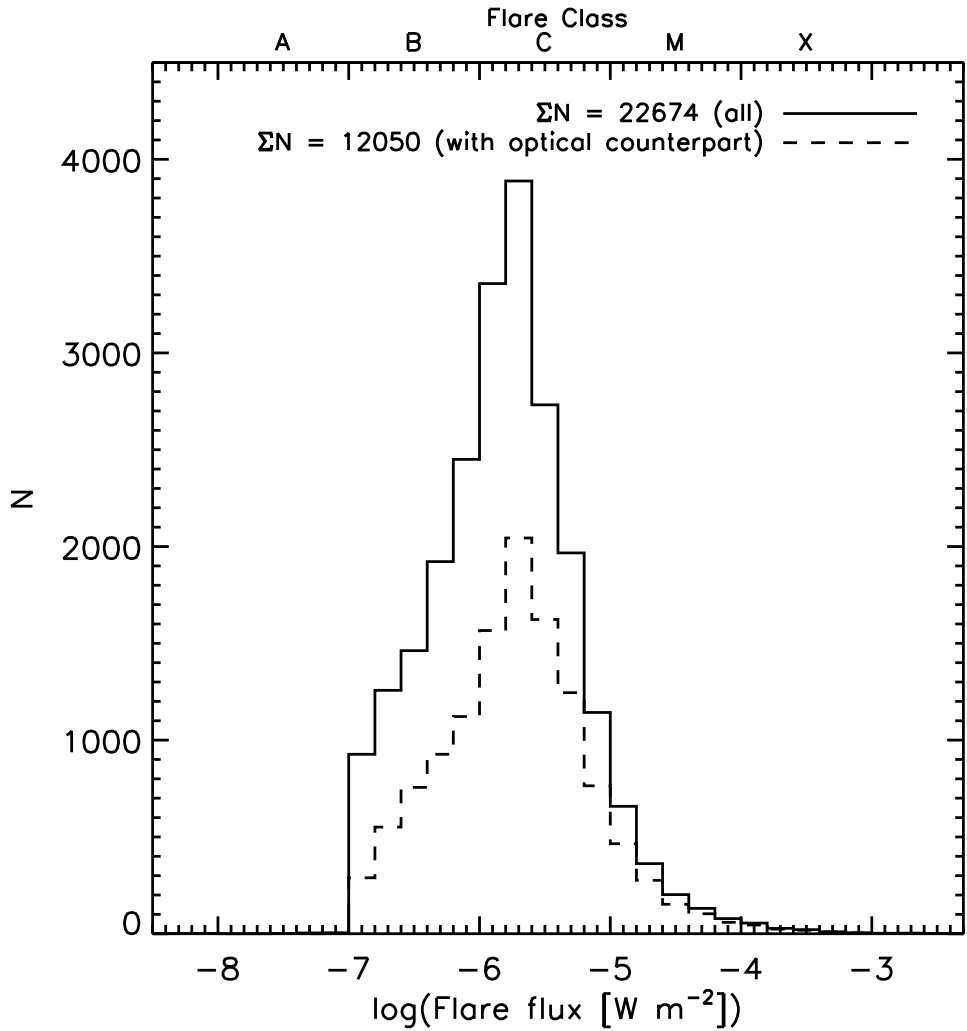}}
\caption{Summary of flare fluxes recorded in the GOES X-ray flare database from 
1996-2006.  22 674 flares in total were recorded, 12 050 of which had measured positions 
of optical counterparts to the X-ray flare.  For correlation with CMEs, we use only the 
12 050 flares with known positions.
}
\label{F-allflares}
\end{figure}

Spanning more than three decades of observation, the Geostationary Operational Environmental 
Satellite (GOES) flare database\footnote{\url{http://www.ngdc.noaa.gov/stp/SOLAR/ftpsolarflares.html}} 
reports the 1-8 \AA\ band full-disk X-ray flux at Earth; 
flare classifications are then applied based upon the peak flux in that bandpass. From 1996-2006, 
the database contains information for 22 674 flares.  Of these, positions for optical counterparts 
are documented for approximately half of the flares in the database (Figure~\ref{F-allflares}).  
Of these 12 050 flares, one A class, 3 638 B, 7 248 C, 1 056 M, and 107 X class flares are recorded.

\section{Determining Flare-CME Association\label{S-statcor}}

As described above, we use the 6 733 CMEs from the LASCO CME catalog with well measured masses and 
the 12 050 flares with optical counterpart positions from the GOES flare database.

\begin{figure}
\centerline{\includegraphics[width=5.5in]{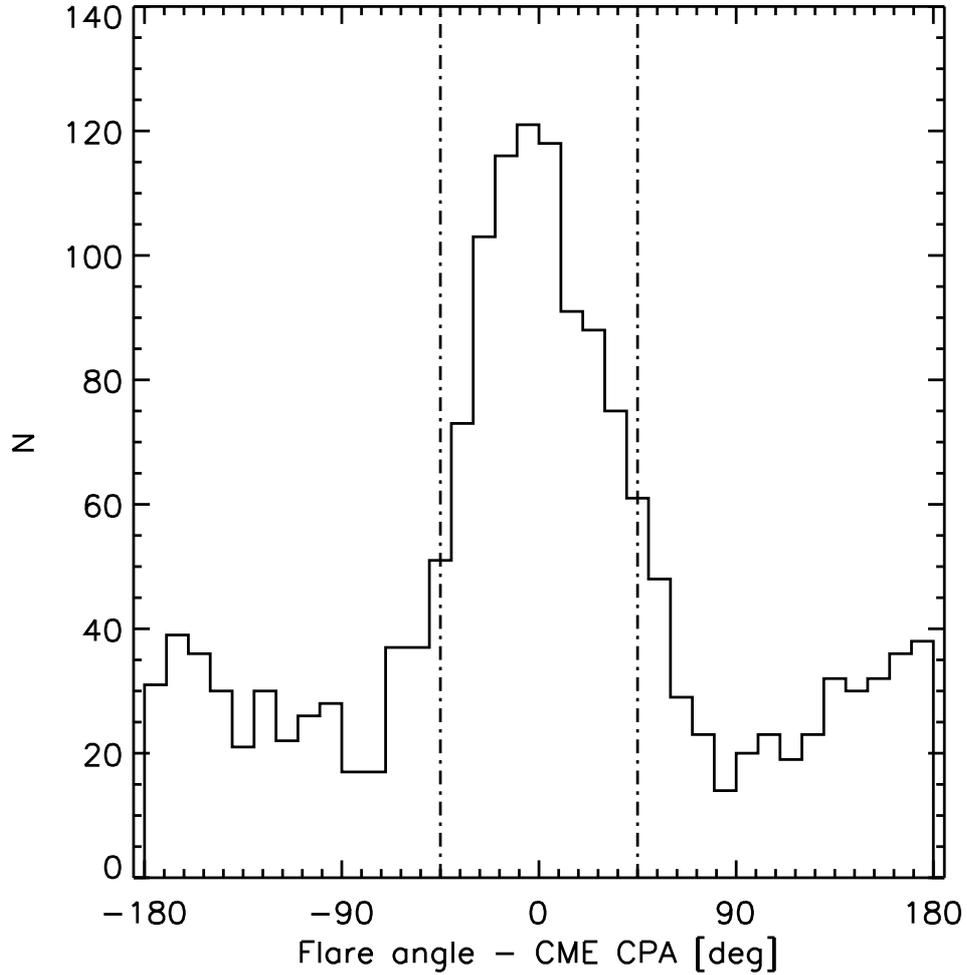}}
\caption{For flares and CMEs matched within a $\pm$2 h time window, the 
angular separation of flares and CME central position angles shows a clear peak 
about 0\degr$\pm$45\degr (vertical lines).
}
\label{F-angularseps}
\end{figure}

In determining whether a given flare and CME are associated, we utilize spatial and temporal 
criteria, requiring that both flare and CME occur within a set time window and angular separation.
We initially set the time window generously to select CMEs which occur within $\pm$2 h of a 
flare's start time (though we refine this below).  Converting Stonyhurst system flare positions 
(Cartesian) on the disk to spherical 
coordinates, we show in Figure \ref{F-angularseps} the angular separations of these time-matched CMEs 
and flares. A clear peak is seen about 0\degr$\pm \approx$~70\degr; we adopt a separation 
criterion of $\pm$45\degr\ (\textit{e.g.}, both events occur in the same quadrant of the disk with respect 
to the CPA of the CME) as a value intermediate to the $\pm$70\degr\ in our Figure \ref{F-angularseps} 
and the $\approx\pm$30\degr\ results of \inlinecite{Yashiro:2008}. Interestingly, but beyond the 
aims of this work, we note that there is a significant peak for flares which occur $\approx$180\degr\ 
in separation from the CME CPA, potentially indicative of large-scale correlated disruptions.

\begin{figure}
\centerline{\includegraphics[width=5.5in]{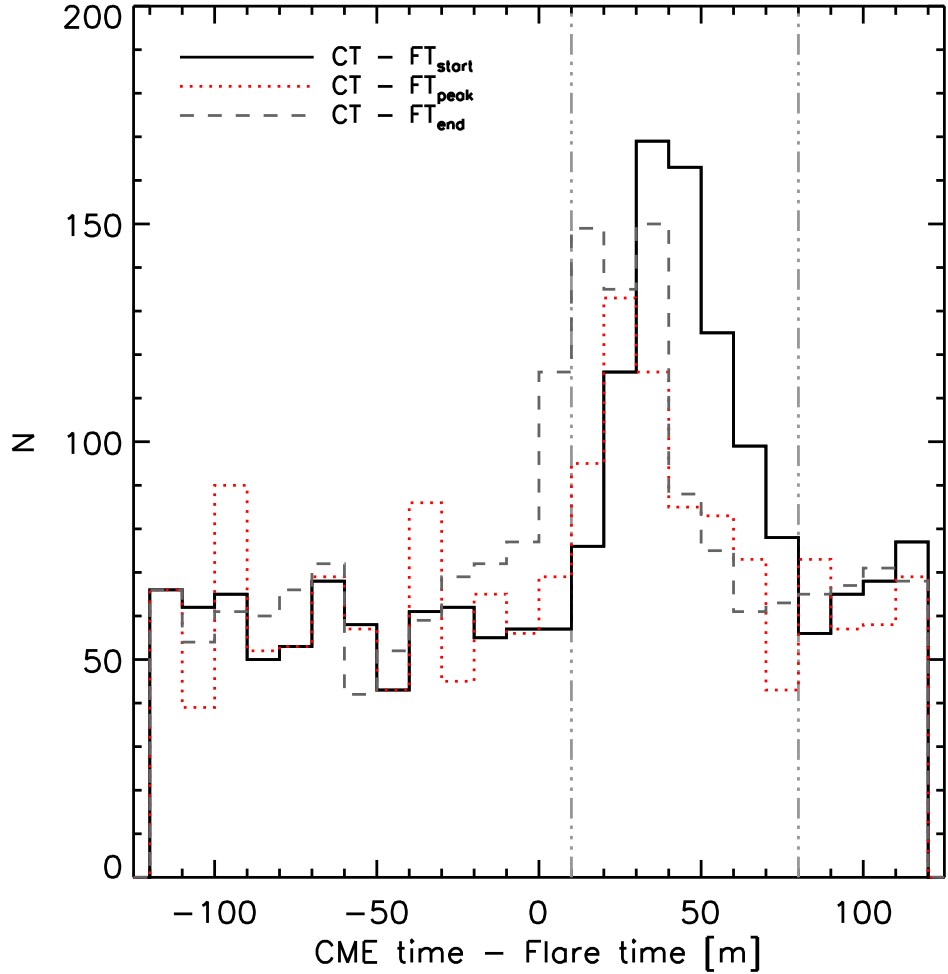}}
\caption{Assessment of which flare time to use for associating flares and CMEs; 
CT and FT refer to CME start time and flare time, respectively.  
We show here a $\pm$2 h time constraint as well as the $\pm$45\degr\ CME CPA-flare 
angle separation applied, the difference in each histogram being which flare 
time was used. Previous studies (\textit{e.g.}, Mahrous \textit{et al.}, 2009) found degree of 
flare-CME association to be dependent upon whether flare start, peak, or end time was 
used in flare-CME pairing.  Time offsets are shown here of CME start time minus flare 
start (solid, black), peak (dashed, red), and end times (dot-dashed, gray).  We observe 
the strongest peak in number of pairs when using the flare start time. The peak of the 
black histogram is a factor of $\approx$3 above the ``background'' and well-defined in shape, 
spanning a range of time offsets from 10-80 min (triple dot-dashed, light gray 
vertical lines).
}
\label{F-tdiffs}
\end{figure}

After applying the $\pm$2 h temporal and a $\pm$45\degr\ angular separation cut, we assess 
whether the initially chosen, large time window can be narrowed (Figure \ref{F-tdiffs}). 
Regardless of which flare time (start, peak or end) we use, there is a significant peak 
between 0-80 min in the resulting flare/CME pairs' time separations (CME start time - 
flare time).  Other time separations surrounding these peaks appear to represent a ``background'' 
level at $N\approx$60 pairs per 10 min time bin; these could represent randomly matched 
flare-CME pairs which are not truly associated.  Events appear to be the most strongly correlated 
when using the flare start time  (solid, black histogram in Figure~\ref{F-tdiffs}); this is to say 
that the number of events in the 10-80 min bins are greater in number and there is less 
``background'' variation than is seen for the other flare time choices. We therefore adopt the 
flare start time, but in any case this choice does not strongly affect our final results.

It is important to note that the CME ``start time'' in fact corresponds not to the time when the 
CME was launched, but rather to the time the CME first appears in the C2 detector (at a height 
of $\approx$3.23 $R_\odot$ on average). Using the linear speeds of the CMEs reported in the LASCO 
database, we can calculate the approximate CME travel times from the solar surface. The average 
travel time so computed is $75 \pm 17$ min, corresponding very well to the average time offset 
between the CME ``start times'' (detection times) and their associated flare times 
(Figure \ref{F-tdiffs}). Note that this simple calculation assumes negligible acceleration of the 
CMEs. Thus, while we cannot say with certainty that any one correlated flare/CME pair does not 
truly have a time delay, the data are broadly consistent with flare/CME start times that are 
simultaneous at the solar surface.

Interpreting the 10-80 min peak of Figure \ref{F-tdiffs} as a time offset region during which there 
is a higher probability of finding associated flares and CMEs, we narrow the time correlation window 
to accept CMEs which occur 10-80 min after a flare (vertical, triple dot-dashed lines, Figure 
\ref{F-tdiffs}). This is consistent with the findings of \citeauthor{Andrews:2001} (\citeyear{Andrews:2001}), 
\citeauthor{Mahrous:2009} (\citeyear{Mahrous:2009}); in the latter work, the analysis used 
flare peak flux times and end times. This 10-80 min choice in temporal separation could potentially 
eliminate some true flare-CME pairs---previous studies (\textit{e.g.}, \citeauthor{Harrison:1991}, 
\citeyear{Harrison:1991}, \citeauthor{Harrison:1995}, \citeyear{Harrison:1995}) have shown instances 
in which CMEs precede flares---but we do not consider these cases due to the absence of a distinct 
peak in the range of negative time offsets. The 10-80 min window also evidently includes some 
``background'' events. Potentially, these cases only represent a minority of flare-associated CMEs.

Since we cannot tell from this data set whether a CME originates on the visible side of the limb 
or just beyond it, it is possible that some ``back-sided'' CMEs are spuriously associated with 
flares on the visible side. We can estimate the effect of back-sided CMEs using geometric arguments. 
First, we assume that CMEs observed by LASCO can be detected if they originate from a 260\degr\ 
longitudinal range about the Sun (\textit{i.e.}, from the full 180\degr\ of the visible disk, plus 40\degr\ 
beyond both the eastern and western limbs), and we assume that CMEs are equally probable to 
originate from anywhere within this area. Thus, 33\% of all CMEs (80\degr$/$240\degr) could be 
back-sided CMEs that are then potentially paired with flares on the visible solar disk. Our spatial 
cut requires $\pm$45\degr\ coincidence in polar angle; assuming CMEs are equally likely to occur 
over all 180\degr\ of polar angle, this serves to eliminate half of all 
spurious correlations, reducing the number of back-sided CMEs potentially introduced to our 
correlated sample to 16\%. Further reducing this is our time-matching requirement, wherein we select 
only the CME-flare pairs occurring within 70 min, out of the full 240-min time window searched 
(\textit{i.e.}, 70/240 = 29\%). Combining these, we potentially include 5\% of back-sided CMEs in our flare-CME 
pairs. Of course, some of these back-sided CMEs may in fact be truly associated with the observed 
flares; however, as a conservative statement, back-sided CMEs should represent at most a 5\% 
contribution of spuriously flare-correlated CMEs in our sample and thus likely add only modestly to 
the scatter in our correlated quantities.

\section{Results\label{results}}

\subsection{Frequency of Associated Flares and CMEs}\label{results.1}

\begin{figure}
\centerline{\includegraphics[width=5.5in]{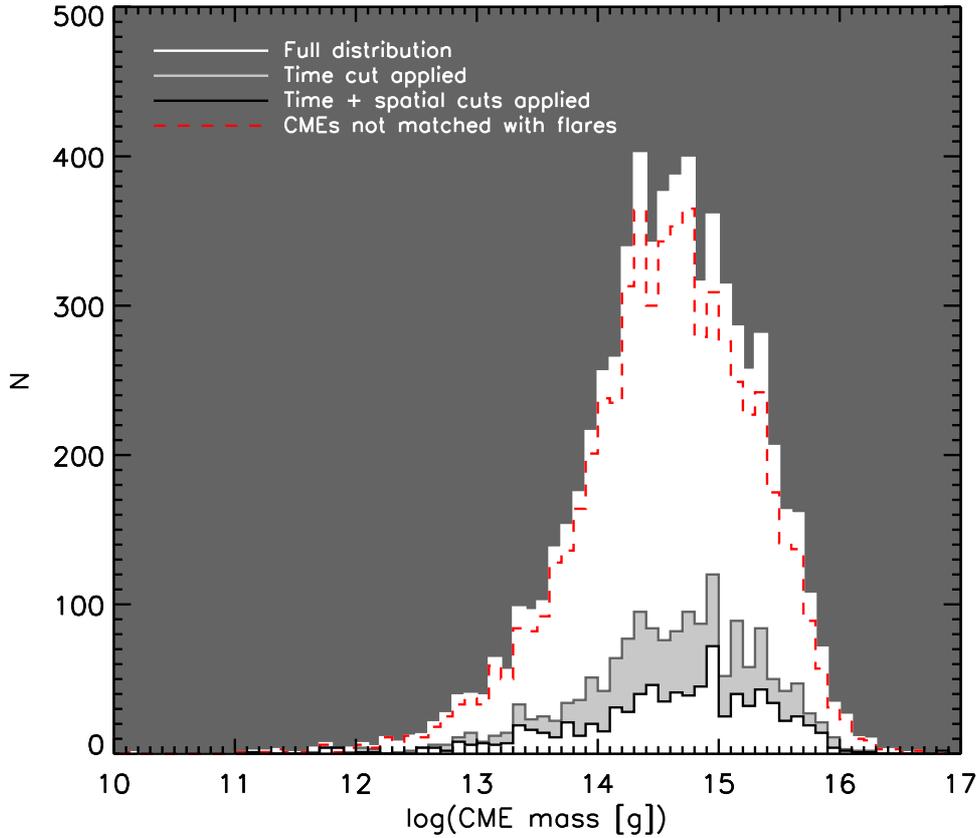}}
\caption{Change in the number and distribution of CME masses as flare/CME pairs are correlated.
The white histogram bounded by dark gray shows the full initial distribution of well 
constrained CME mass measurements. Pairing flares with CMEs occurring 10-80 min 
after the flare start time, the data set is greatly reduced in number (filled, light gray 
histogram).  The position criterion, CME and flare position angle equality within 
$\pm$45\degr, leaves 826 flare/CME pairs (solid black line, white filled histogram).  
In red, we show CMEs not associated with flares by these criteria.
}
\label{F-cmemasscuts}
\end{figure}

With each of the above constraints applied, the total number of flare-CME pairs in the sample is winnowed 
as illustrated in Figure~\ref{F-cmemasscuts}. The final sample contains 826 flare-CME pairs, 737 of 
these are unique. As performed, the correlation allows for the selection of multiple CMEs per 
flare, and one CME may be found to be associated with multiple flares. With the data available, there 
is no way to distinguish which CME may ``actually'' be associated with the flare, or if indeed there 
should be only one associated. We therefore use all 826 pairs when comparing flare and CME properties, but
we use only the 737 unique CMEs' properties when analyzing properties of flare-associated CMEs vs 
CMEs occurring without flares.

Of the 6 733 CMEs we perform the correlation with, 737 unique CMEs are found to be associated 
with flares; put another way, we find 11\% of CMEs to be flare-associated.
Here we assess whether our applied criteria for correlation or observational 
effects could be decreasing the number of associations found, and in which variables these issues
may create the most problems.  In particular, the LASCO data gaps in addition to CME width and 
projection effects likely make our 11\% an underestimate. 
For comparison, previous estimates of CMEs associated with flares are 40\%, 
\cite{Munro:1979} and 34\% \cite{StCyr:1991}.

\subsubsection{Data Gaps}

LASCO coverage was 83\% complete from 1996-2006. We found that the following numbers of flares occurred 
during a LASCO data gap: 24 X, 119 M, 595 C, and 803 B class. These flares could have been ``lost'' 
to our analysis, as the data gaps ended 10 min or more after the flare, within our 10-80 
min temporal selection window.  These values do not reflect, however, cases in which a CME 
did occur after the data gap ended and the flare was paired with that CME; we could not exclude 
these cases from the analysis as the technique does not allow for discerning which flares and 
CMEs are ``actually'' associated.  Proportionately, more 
high energy flares occur during data gaps, likely because the particle events associated with the 
flares cause gaps in coverage by the LASCO C2 detector. B flares are excluded from this statement, as their 
energies are near the detection threshold for the GOES detector. An additional caveat to this 
is that the numbers of ``lost'' flares reported above do not account for simultaneity of flares; 
multiple flares could occur in a given data gap (thus, we cannot conclusively link a given class 
flare to a higher frequency of particle events), and we have counted each singly. 

\subsubsection{Wide, Partial Halo, and Halo CMEs{\label{S-derive_UB}}}

Surprisingly, we find that in the 
application of our time and angular separation constraints, most of the 107 X class flares 
are lost from the sample. Only seven of the 826 pairs have X class flares. This contradicts the 
results of \citeauthor{Andrews:2003} \citeyear{Andrews:2003}, who found an almost 100\% X-flare-CME 
association\footnote{Note, however, that \citeauthor{Andrews:2003} \citeyear{Andrews:2003} set 
out specifically to define a sample of associated ``big flares'' and CMEs.}.

\begin{figure}
\centerline{\includegraphics[width=5.5in]{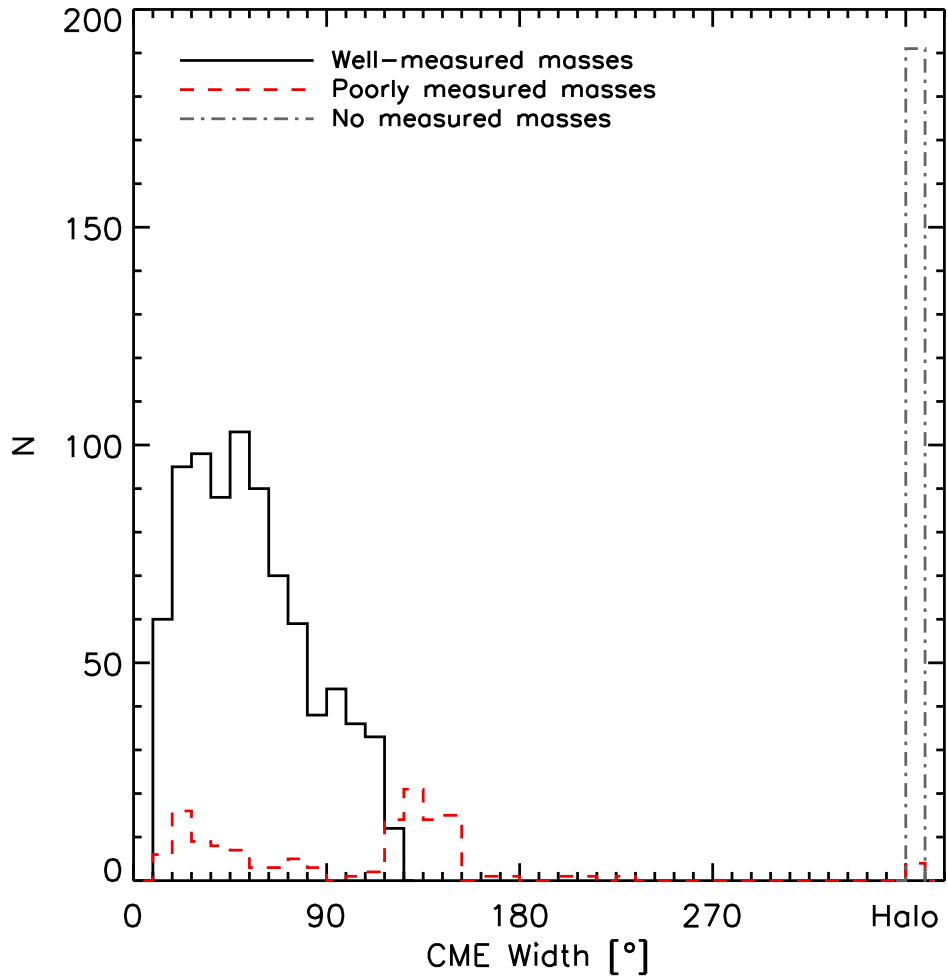}}
\caption{CME widths and frequency of ``halo'' CMEs. The distributions of widths for CMEs 
with well measured masses (solid, black line) and poorly constrained masses (dashed, red 
line) are shown. Masses are more difficult to measure, and thus the most poorly constrained, 
for the widest CMEs. All but three halo CMEs in our sample have no measured masses (dot-dashed, 
gray line).
}
\label{F-cmewidths2}
\end{figure}

A limitation imposed by our data completeness requirement for the flares and CMEs analyzed in this 
work (see Section \ref{data}; we require flares to have positions, and CMEs to have well measured 
masses) is that we have excluded halo CMEs, as their masses cannot be 
measured via the same technique used for CMEs perpendicular to our line of sight. 
Additionally, our analysis excluded CMEs which had high errors in their mass measurements; these 
CMEs are illustrated in Figure \ref{F-cmewidths2} as a red histogram. Most of the CMEs with poorly 
constrained masses are what would be designated as ``wide'' or ``partial halo'' CMEs (\textit{i.e.}, 
with observed widths greater than 60\degr).

In an effort to account for these selection effects in our analysis, we performed the correlation again, 
including wide, partial halo, and halo CMEs. This inclusion increased the CME sample size to 8 128 
CMEs. Because they lack clearly defined CPAs, we did not
require spatial coincidence for halo CMEs and flares, thereby including any halo CMEs which were 
time-correlated with flares. CMEs with poor mass measurements typically do have documented CPAs, so 
those events were required to be co-spatial with associated flares within $\pm$45\degr, as previously.  
This resulted in 1 153 flare-CME pairs, with 53 X, 278 M, 608 C, and 214 B class flares. 
Of the 1 153 pairs, 1 028 of the CMEs are unique, resulting in a 13\% flare-association; this is 
effectively indistinguishable from the association fraction observed when including only CMEs with 
good masses (Section \ref{results.1}).
While the entire sample is increased by $\approx$40\% and we don't find a greater fraction of CMEs to 
be associated with flares, 
the greatest percentage increases occur with higher flux flares. Including CMEs with poorly measured 
masses and halo CMEs results in $\approx$8 times as many X class flares, and almost double the M class 
flares. C and B class flares show only modest $\approx$20\% increases in representation in the sample.
We assess below (Section \ref{S-massflux}) the main effect this could have on our results, and show a 
method to possibly correct for it.

In summary, we lose $\approx$75\% of X class flares to the combination of data gaps and observational 
selection effects, and this ``loss fraction'' is greatest for the strongest flares, decreasing with 
decreasing flare flux.
Despite these factors potentially over-narrowing our correlated sample, 
we have demonstrated that the requisite time and position conditions applied here are sound in 
reasoning and represent a conservative approach which may only under-count flare-CME pairs.

\subsection{Correlations in the Matched Flare-CME Sample}

For the matched sample of 826 flare-CME pairs (see Section \ref{S-statcor}), we assess potential 
relationships between five different flare and CME parameters: flare flux, CME linear speed, CME 
acceleration, CME width, and CME mass.  Primarily, we seek to define a flare flux-CME mass 
relationship. We also perform additional correlations with our large flare-CME pair sample in 
order to compare with previous work and to elucidate the underlying physics.

\subsubsection{CME Linear Speed, Acceleration, and Flare Association\label{S-flarecmespeed}}

\begin{figure}
\centerline{\includegraphics[width=5.5in]{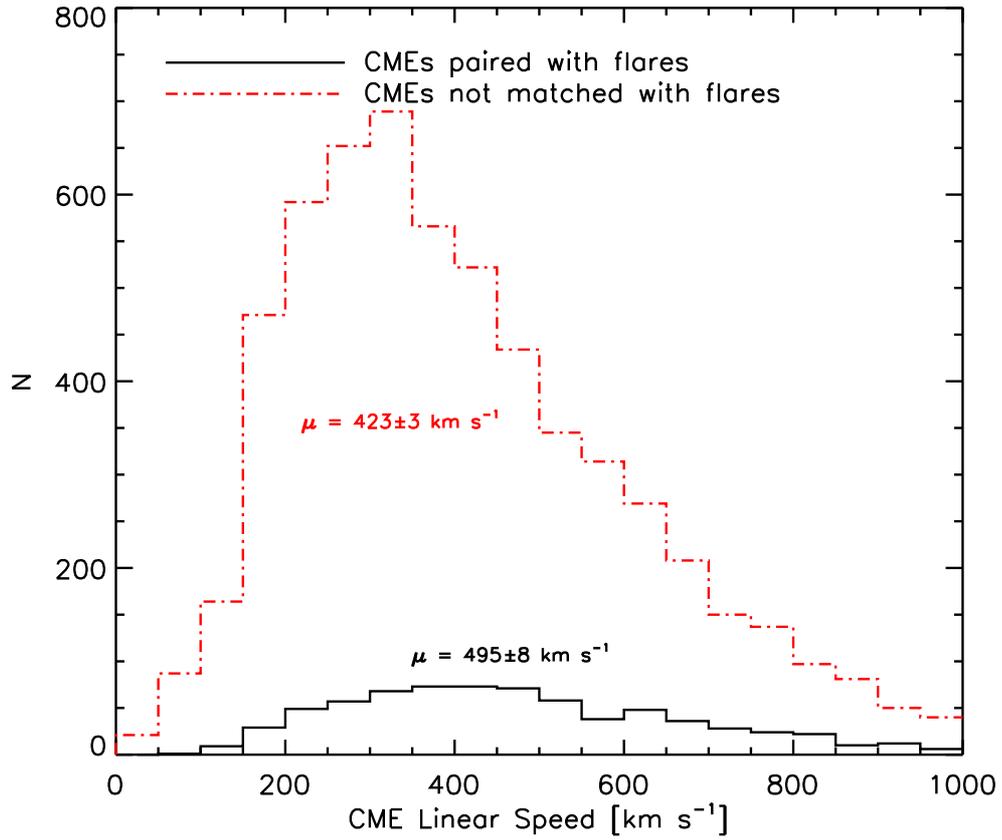}}
\caption{As in Figure~\ref{F-cmemasscuts}, the red and black distributions show CMEs 
not matched with flares and CME/flare pairs, respectively. We find that CMEs 
associated with flares have, on average, faster linear speeds (see Section \ref{S-flarecmespeed}).
In this and subsequent figures, we report on the plot the mean of each distribution 
with the error on that mean (\textit{i.e.} the standard deviation of the mean, $\sigma$/$\surd$N).
}
\label{F-cmelinspeeds}
\end{figure}

Flare and eruptive prominence-associated CMEs have been observed to have higher linear speeds than 
those which are not associated with flares \cite{Gosling:1976,Moon:2003}.  To check whether our 
flare-CME pairs reflect this observed relationship, we show in Figure~\ref{F-cmelinspeeds} the 
two distributions of CME linear speeds for our sample. The mean linear speed for CMEs associated 
with flares is 495$\pm$8 km s$^{-1}$; CMEs not associated with flares have an average speed of 
422$\pm$3 km s$^{-1}$ (the error we report here is the standard deviation of the mean).  A $t$-test 
\cite{Press:1995} assesses the probability that two distributions have significantly different 
means. The $t$ statistic for these distributions is 8.5, and the corresponding null-hypothesis 
probability (that the two distributions have the same mean) is $\ll$10$^{-6}$.

Due to the asymmetric skew of the distributions, we verify this result by performing a 
non-parametric analog of the $t$-test, the Wilcoxon-Mann-Whitney test (also known as a $U$ test). The 
statistic is Z, and the corresponding probability assesses the likelihood, when sampling two 
distributions, of preferentially finding a higher value in one distribution than another.  The 
result of the $U$ test corroborates the $t$-test: the $Z$ statistic is -9.0, and the null-hypothesis 
probability is $\ll$10$^{-6}$. We conclude that these two distributions have statistically 
significantly different means, consistent with observations that flare-associated CMEs have higher 
linear speeds than CMEs not associated with flares.

\begin{figure}
\centerline{\includegraphics[width=5.5in]{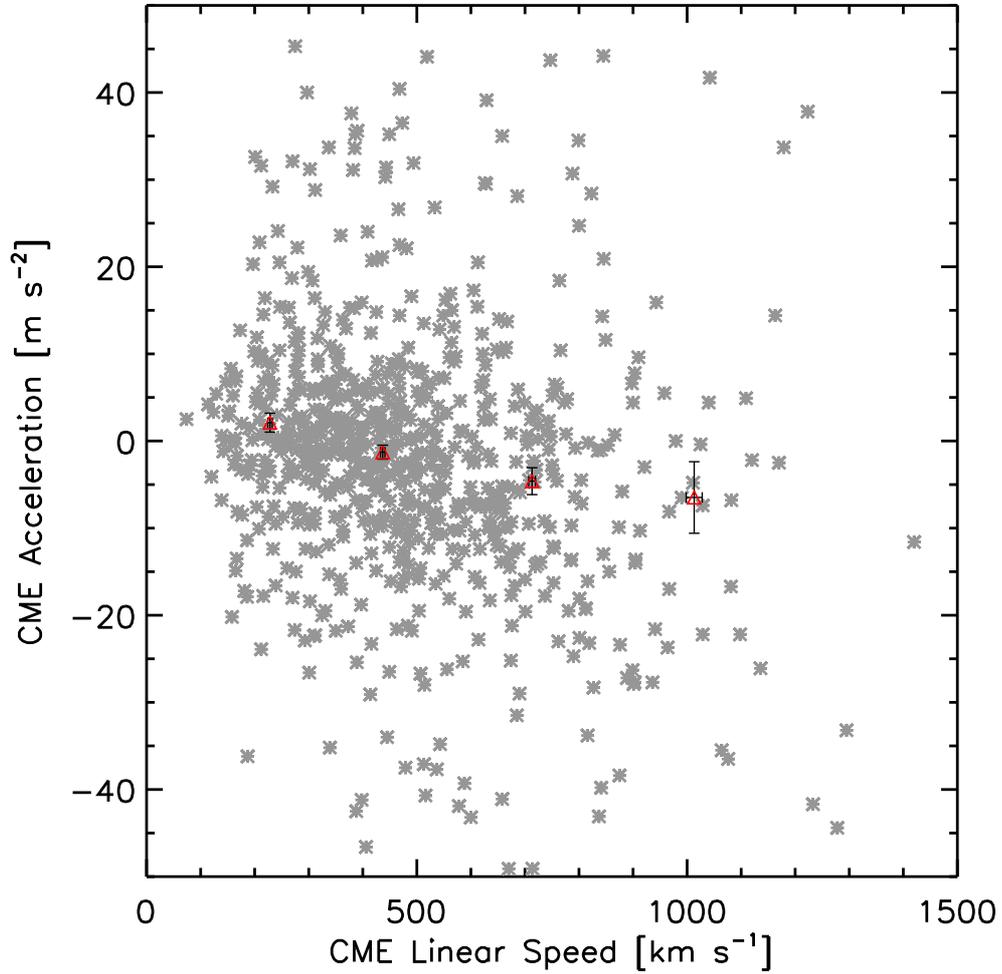}}
\caption{For flare-associated CMEs, linear speeds and accelerations are shown as 
gray asterisks. In bins 300 km s$^{-1}$ wide in linear speed, mean speeds and accelerations 
are plotted as triangles (red). The figure has been cropped so the y-axis range of the 
triangles (red) can be clearly seen; 13 data points are not shown, lying beyond $\pm$50 m s$^{-2}$.
While the magnitude of the slope of these points is rather small, its existence is 
significant: a Spearman's $\rho$ correlation test shows these two quantities are highly 
significantly anticorrelated ($\rho =$ -0.22, null-hypothesis probability $\ll$10$^{-6}$).
We see here that the fastest CMEs tend to decelerate, while the slower CMEs accelerate 
through the C2 field of view (Section \ref{S-flarecmespeed}).
} 
\label{F-speedaccel}
\end{figure}

Figure \ref{F-speedaccel} shows the relationship between CME speed and acceleration. Evidently, 
there is a trend of decreasing acceleration (or rather, increasing deceleration since the mean 
accelerations are negative) with increasing 
CME linear speed. To test this statistically, we perform a Spearman correlation test, finding 
the $\rho$ coefficient to be $-$0.22---a monotonically decreasing relationship---and the corresponding 
null-hypothesis probability $\ll$10$^{-6}$.  In summary, we see that the fastest CMEs are associated 
with flares, and the fastest CMEs decelerate the most on average.

\subsubsection{CME Mass, Acceleration, Flare Flux\label{S-cmassaccelfflux}}

\begin{figure}
\centerline{\includegraphics[width=5.5in]{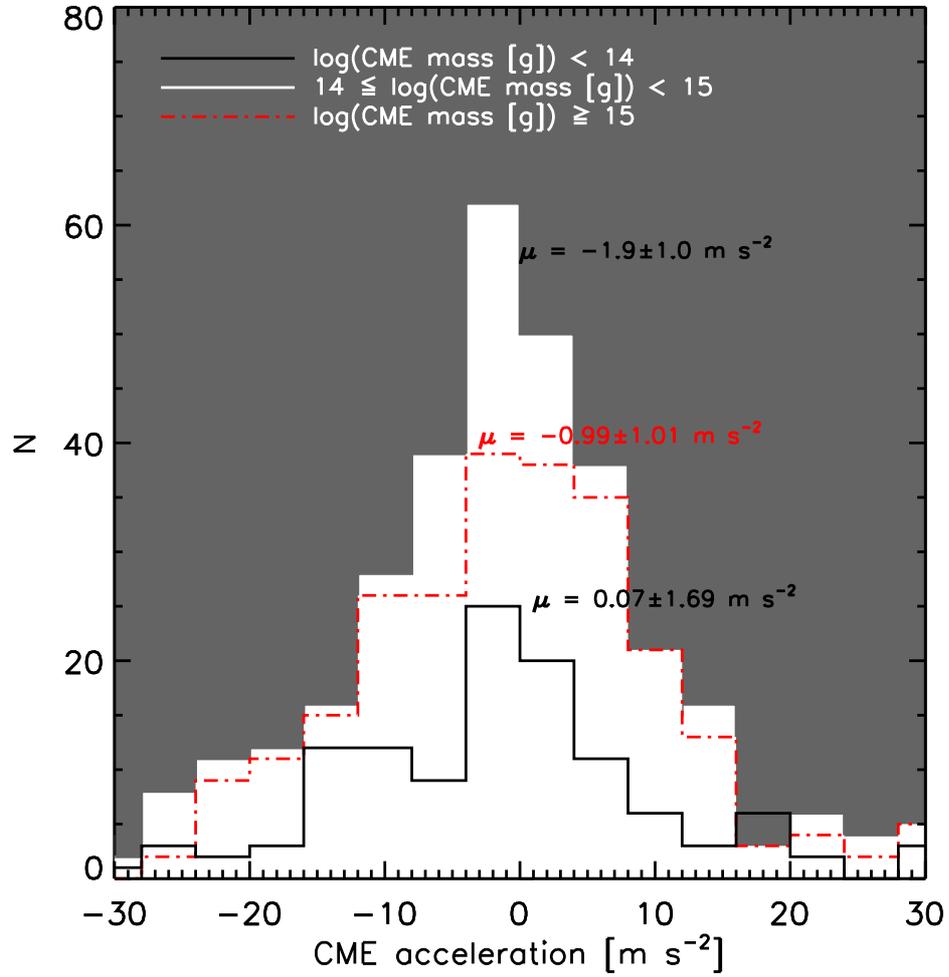}}
\caption{Relationship between CME mass and CME acceleration. The dot-dashed (red) 
histogram represents accelerations for the highest mass CMEs, with log(CME mass [g]) 
$\geq$ 15. The white histogram bounded by dark gray shows accelerations for CMEs with 
masses $\geq$ 10$^{14}$ g and $<$10$^{15}$ g. The lowest masses shown, $<$10$^{14}$ g, are 
represented by the solid (black) outlined histogram. There is no apparent dependence of 
CME acceleration on CME mass (see Section \ref{S-cmassaccelfflux}).
}
\label{F-cmeaccels2}
\end{figure}

In Figure \ref{F-cmeaccels2} we examine more closely the potential effect of CME mass on acceleration 
for CMEs associated with 
flares. The highest mass division includes CME masses $\geq$10$^{15}$ g, 
while the middle and lowest mass subsets span the ranges of 10$^{14}$ g $\leq$ CME mass 
$<$ 10$^{15}$ g and CME mass $<$ 10$^{14}$ g, respectively. Again utilizing the $U$ test, we 
find that all three distributions have statistically indistinguishable means.

\begin{figure}
\centerline{\includegraphics[width=5.5in]{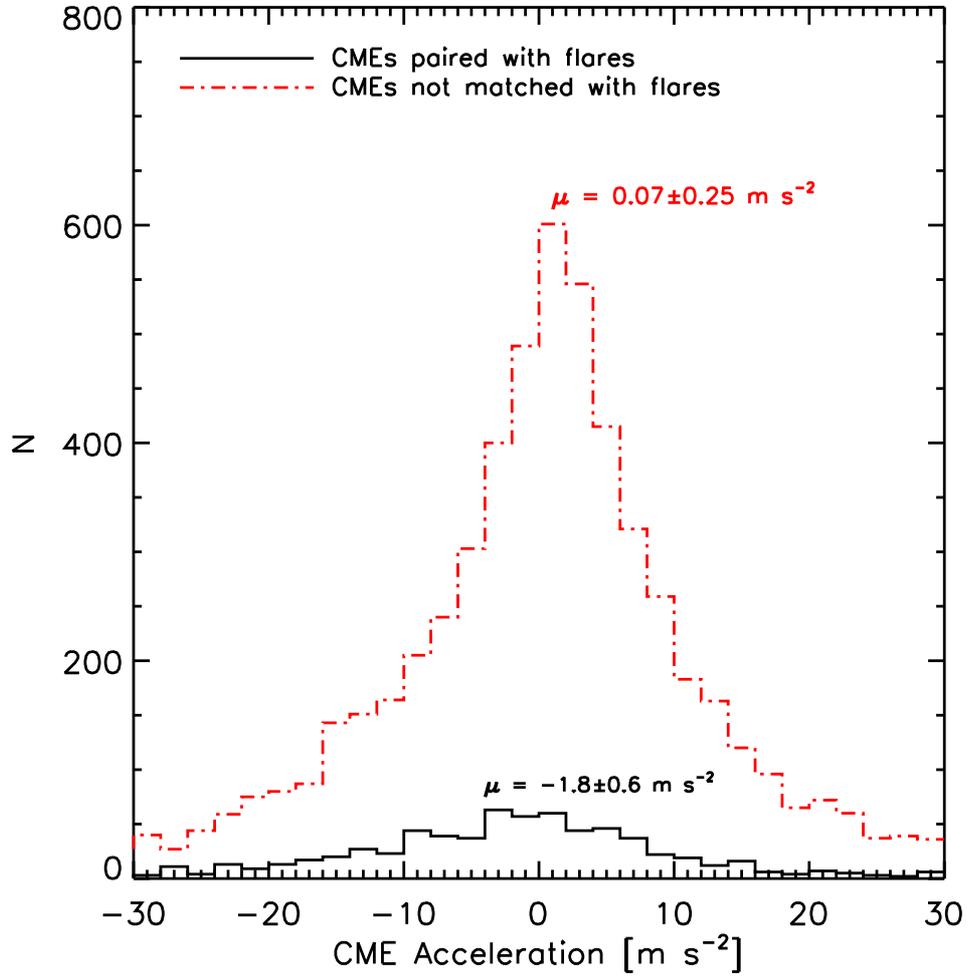}}
\caption{CME accelerations: CME/flare pairs (solid, black distribution) and CMEs not associated 
with flares (dot-dashed, red). The means of the two distributions (printed on the plot, with the 
associated standard deviations of each mean) are similar in magnitude 
but significantly different. The $t$ statistic is -2.5, with a null-hypothesis 
probability of 0.01. A non-parametric $U$ test confirms this, with a $Z$ statistic of 4.6 and a 
null-hypothesis probability of 2$\times$10$^{-6}$ (see Section \ref{S-cmassaccelfflux}).
}
\label{F-cmeaccels1}
\end{figure}

In Figure~\ref{F-cmeaccels1}, CME acceleration in the flare-associated case and the converse 
scenario is shown.  The mean of the CME-flare pair distribution is -1.8$\pm$0.1 m s$^{-2}$, while 
the mean of the distribution of CMEs without flares is 0.07$\pm$0.25 m s$^{-2}$ (the errors on 
these quantities are the standard deviations of the mean). 
We use a $t$-test to determine whether the acceleration distributions 
have significantly different means.  The $t$ statistic is -2.5, the corresponding null-hypothesis 
probability 0.01;  
this indicates these distributions do indeed have significantly different means. A 
$U$ test was also performed; the $Z$ statistic found to be 4.6, and the null-hypothesis probability 
is 1.9$\times$10$^{-6}$.  The CME-flare pairs have CMEs which tend to decelerate on average, 
while the CMEs not associated with flares are centered about 0 acceleration.

\begin{figure}
\centerline{\includegraphics[width=5.5in]{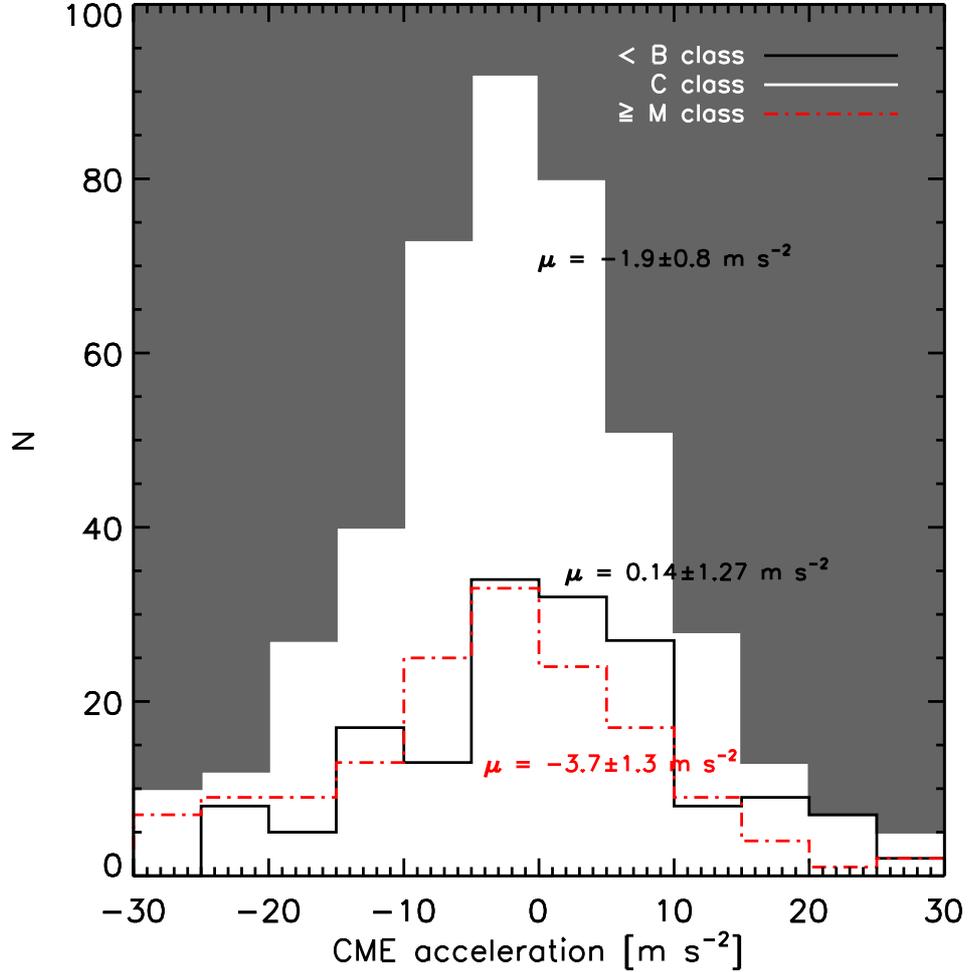}}
\caption{CME acceleration binned by flare flux. M and X class flares (log(flux [W m$^{-2}$]) 
$\geq$ -5; combined here due to small number of X class flares in sample) are paired with 
CMEs that have accelerations as shown by the dot-dashed (red) distribution. C class flares 
(-6 $\leq$ log(flux [W m$^{-2}$]) $<$ -5) have corresponding CME accelerations as shown by 
the white histogram bounded by dark gray. A and B class (log(flux [W m$^{-2}$]) $<$ -6) 
flares' associated CMEs have accelerations as shown by the solid (black) histogram. We report 
the mean value of each distribution, with the standard deviation of each mean, on the figure.
The B 
class flare/CME acceleration distribution has a mean significantly different than the higher 
energy flare associated CMEs; CMEs associated with lower energy flares appear to modestly 
accelerate, while those associated with higher energy flares decelerate (see Section 
\ref{S-cmassaccelfflux}).
}
\label{F-fluxaccels}
\end{figure}

We next assess how the magnitude of this 
deceleration for flare-associated CMEs may relate to the flux of the associated flare.
In Figure~\ref{F-fluxaccels}, the relationship between CME acceleration and flare class is shown 
for the 826 CME-flare pairs. X and M flares are combined for this comparison, as there are only 7 
X class flares in the 826 flare-CME paired sample. We employ the $U$ test, making no assumption that 
these distributions are Gaussian. Comparing the mean accelerations of B class and weaker with C 
class flare-associated CMEs, the 
$Z$ statistic is -1.9, with a corresponding null-hypothesis probability of 0.03. 
Applying the $U$ test to intermediate strength flares, the difference in 
deceleration between C, M, and X class flare-associated CMEs is not statistically significant.
However, testing the same group of B/A class 
flare-associated CMEs against M class and stronger flare-associated CMEs, the $Z$ statistic is -2.8, 
its null-hypothesis probability 0.002. Evidently, just as flare-associated CMEs decelerate on average, CMEs 
associated with the strongest flares show the most deceleration.

\subsubsection{CME Width, Mass, and Flare Flux\label{S-widthflareflux}}

\begin{figure}
\centerline{\includegraphics[width=5.5in]{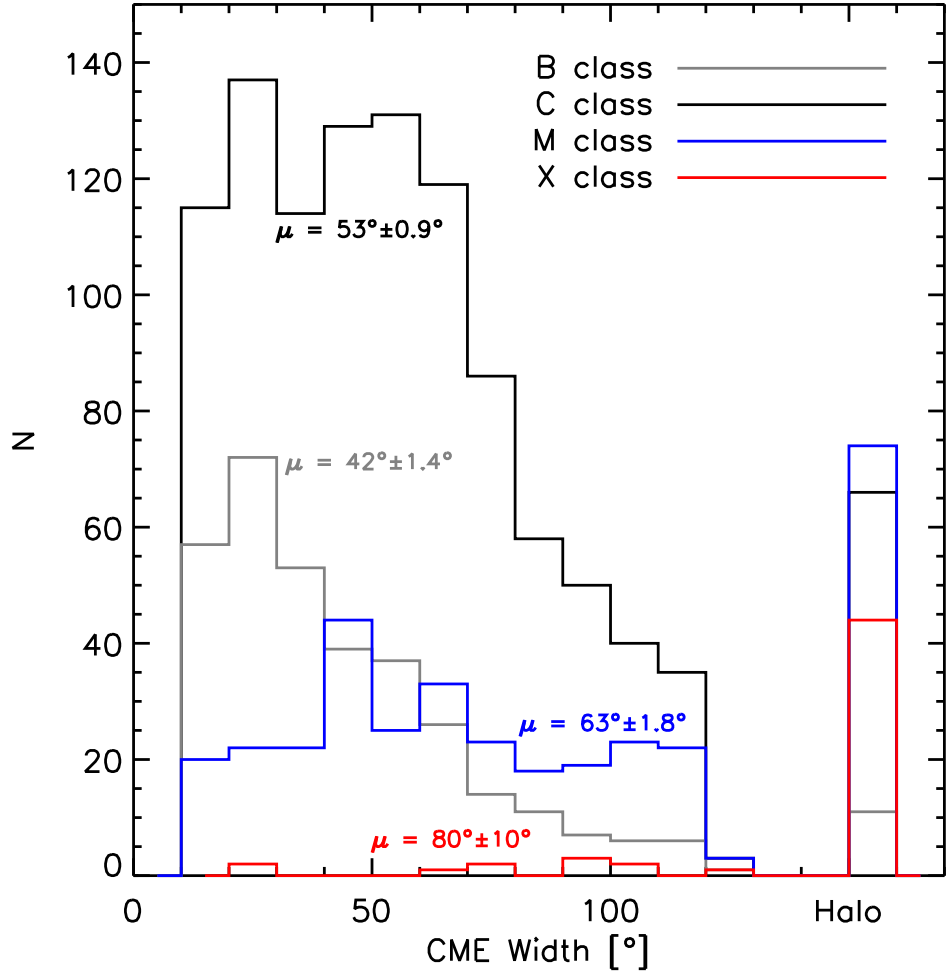}}
\caption{Widths of CMEs associated with flares of a given flux. The mean widths for each flare 
class are printed on the plot with the standard deviations of each mean. 
The mean widths for CMEs associated with M, C, and B class flares are statistically 
significantly different from one another. Interestingly, 
80\% of CMEs associated with X class flares have ``halo'' designations, that percentage 
decreasing with decreasing flare flux (see Section \ref{S-widthflareflux}).
}
\label{F-cmewidths}
\end{figure}

We have so far shown above that higher energy flares are associated with faster CMEs that 
decelerate the most. To test whether higher energy flares are also associated with faster, 
more massive CMEs, we checked to see if the width of the CME was related to the flux of 
the associated flare. In Figure \ref{F-cmewidths}, we show that indeed, higher flux flares are 
associated with wider CMEs. With increasing flare flux, the width of the associated CME increases: 
the mean widths (plus or minus the standard devations of the mean widths) of the flare-associated 
CMEs are 80$\pm$10\degr, 63$\pm$1.8\degr, 53$\pm$0.9\degr, 
and 42$\pm$1.4\degr\ for X, M, C, and B class flares, respectively. 

Considering projection effects, it is unsurprising then that as the mean CME width increases, so too 
does the occurrence of halo CMEs. Halo CME occurrence is also shown in Figure \ref{F-cmewidths}, 
with 80\% of X class flares being associated with halo CMEs, while M, C, and B class flares have 
halo CME associations of 21\%, 6\%, and 3\%, respectively. A $t$-test shows the mean widths of CMEs 
associated with M, C, and B class flares are significantly different. Comparing the mean widths of 
CMEs associated with M and C class flares, the $t$ statistic is 5.4, its null-hypothesis probability 
$\ll$10$^{-6}$. For the mean widths of CMEs associated with C and B class 
flares, the $t$ statistic is 6.2, and the corresponding probability is also $\ll$10$^{-6}$.

Finally, we find that as CME width increases, so too does the mass of the CME (Figure 
\ref{F-cwidthmass}). Put another way, the widest CMEs aren't necessarily more diffuse than the 
more narrow CMEs.

\subsubsection{CME Mass, Flare Flux\label{S-massflux}}

\begin{figure}
\centerline{\includegraphics[width=5.5in]{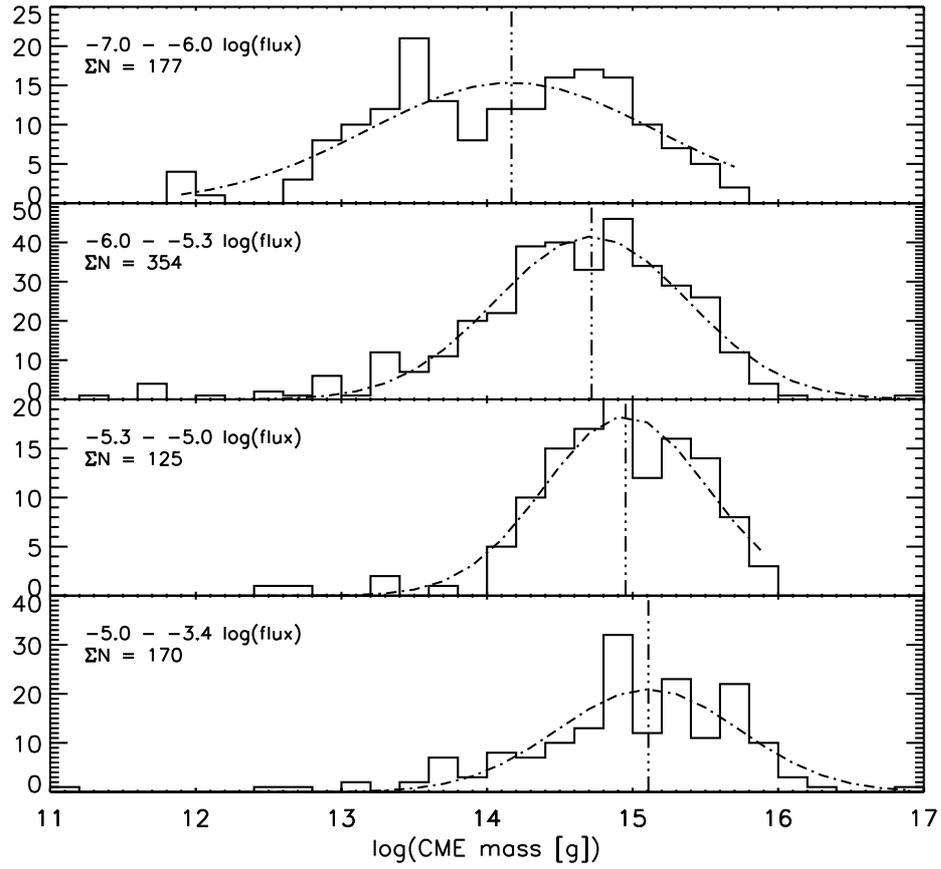}}
\caption{Grouping flare-CME pairs by flare flux, ascending flux from top to bottom, we 
see a clear increase in the centroid value of CME mass (Section \ref{S-massflux}).
}
\label{F-gaussfits}
\end{figure}

For the associated flares and CMEs, the relationship between CME mass and flare flux is shown 
in Figure~\ref{F-gaussfits}. Dividing the pairs by flare flux into four groups, 
we fit the resulting distribution with Gaussian functions. It is clear that the Gaussian 
centroids progress to higher CME mass as the flare flux sampled increases (Figure~\ref{F-gaussfits}).  

In Figure~\ref{F-massfluxrel}, the 826 flare/CME pairs are binned into fourteen equal sets of $N =$~59 
pairs each. This division was chosen to maximize the number of events with which to perform statistics. 
Visually, it is apparent that the mean CME mass in each bin increases with flare flux.  There is an 
apparent ``knee'' to the function around log(flare flux) $\approx$ -5.5, so we use two linear regression 
fits to describe the functional relationship on both sides of this ``knee.'' 
The first function is fit to bins 0-8, and the second to bins 7-13, counting from left to right. 
There is slight overlap in that precisely where the ``knee'' of the function occurs is unclear. 
The first linear fit is of the form: 
\begin{equation}
\rm{log(CME~mass)}=(18.5\pm0.57)+(0.68\pm0.10)\times\rm{log(flare~flux)}.
\end{equation}
The second linear fit follows the form: 
\begin{equation}
\rm{log(CME~mass)}=(16.6\pm1.30)+(0.33\pm0.26)\times\rm{log(flare~flux)}.
\end{equation}

We utilize Spearman's $\rho$ correlation test to assess the degree of correlation of flare flux 
and CME mass. The test does not assume a particular form of correlation, rather, a $\rho$ 
coefficient of unity indicates complete monotonic relationship of the variables tested. For 
the mean values in each bin, we find the $\rho$ coefficient to be 0.99 with a null-hypothesis 
probability of $\ll$10$^{-6}$. This $\rho$ coefficient nearly equal to one and the corresponding 
very low probability confirm that these are highly significantly correlated properties.  

We also performed a Pearson linear correlation test. The Pearson coefficient ranges from $-$1 to 
$+$1 and evaluates how linearly related two parameter sets are. We obtain a value of r $=$ 0.96 
when comparing the mean masses and fluxes in Figure~\ref{F-massfluxrel}.  This strong correlation 
means one could nearly perfectly predict the mean behavior, but do note that there is a scatter 
about each mean mass of $\approx$1 dex.

To assess whether a broken fit is indeed necessary, we fit the correlated flux and mass bins of 
Figure~\ref{F-massfluxrel} with a single linear fit. The $\chi^{2}_{\nu}$ value of this fit is 0.998,
while the $\chi^{2}_{\nu}$ values for each of the fits above are 0.75 and 0.76, respectively.
We also determine the robustness of the ``knee'' by re-binning the data and fitting again.  The data 
divided this time into seven bins of $N =$~118, we fit a single line.  This fit has a $\chi^{2}_{\nu}$ of 0.87.
For comparison, we apply another broken 
log-linear fit to the data with the ``knee'' placed at the same point, log(flare flux)$\approx$~-5.5. 
Each of the functions better fits the data ($\chi^{2}_{\nu}$ values of 0.16 and 0.07, 
respectively, potentially indicating we have ``over-fit'' the data).  In both cases,
we find that fitting the CME mass-flare flux relationship with a broken linear function indeed produces 
a better fit than simply fitting all of the data points with a single linear function. We also find that 
the ``knee'' is robust and does not depend upon how the data are binned.

As mentioned in Section \ref{S-derive_UB}, observational effects could be causing ``lost'' flare/CME 
pairs in the highest flare flux regime, which could in turn be causng the knee in the relationship of 
Figure \ref{F-massfluxrel}. In Section \ref{S-derive_UB}, we 
re-introduce wide, partial halo, and halo CMEs to the analysis. Halo CMEs lack mass measurements, but
what we do know about halo CMEs allows us to assign an upper limit to possible masses. We have shown 
that the highest flux flares are most often associated with halo CMEs, and the highest flux flares 
tend to be associated with the widest CMEs (Figure \ref{F-cmewidths}). We have also shown that the widest 
CMEs are the most massive (Figure \ref{F-cwidthmass}).
Putting this expanded sample into a binned representation as in Figure 
\ref{F-massfluxrel}, we obtain a range of CME masses for the given flare flux of the pair, and assign, 
as an upper bound, the maximum possible mass for that bin to the halo CME. In this way, we obtain 
an estimate of the greatest possible mean masses per bin when including halo CMEs.

In Figure \ref{F-massfluxrelUB}, we reproduce Figure \ref{F-massfluxrel}, now including halo CMEs and CMEs 
with poorly constrained masses. The linear fit that describes this relationship is of the form:

\begin{equation}
\rm{log(CME~mass)}=(18.67\pm0.27)+(0.70\pm0.05)\times\rm{log(flare~flux)}.
\end{equation}

In this case, single power-law fit is sufficient to adequately describe the data. As this case assumes 
the maximum possible halo CME mass, the true CME mass/flare flux relationship likely lies in the 
parameter space bounded by the broken fit of Figure \ref{F-massfluxrel} and the single fit shown in 
Figure \ref{F-massfluxrelUB}.

\section{Discussion and Conclusions}

For flares and CMEs arising from the same active regions, one would anticipate some of their properties 
to be correlated.  For example, a less magnetically complex active region may produce lower flux flares 
and lower mass CMEs.  Conversely, a highly magnetically complex active region may produce both 
higher energy flares and more massive CMEs.  

\begin{figure}
\centerline{\includegraphics[width=5.5in]{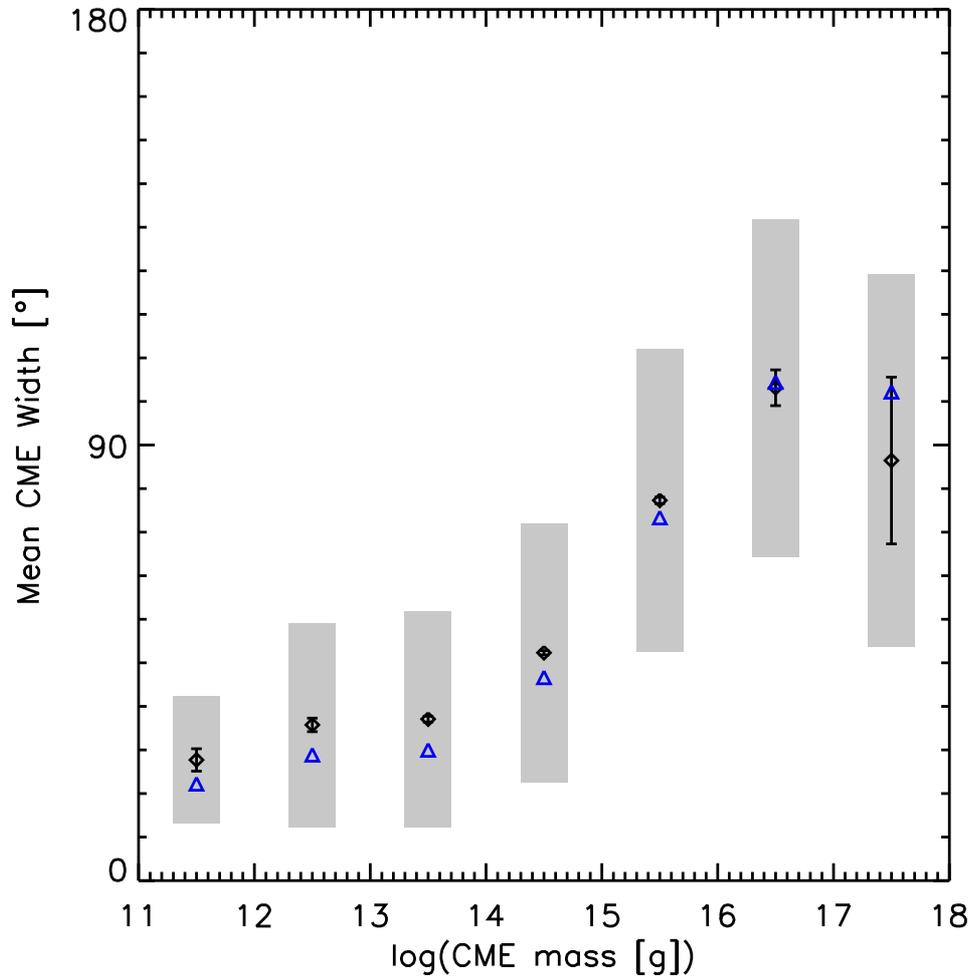}}
\caption{CME width and mass. Each point represents a binned order of magnitude in mass, each 
abscissa value plotted at the bin's center. The diamonds (black) are the mean CME width for 
each mass bin, and triangles (blue) are median values. Error bars show the error on each mean 
value, while the gray shaded boxes show the standard deviation of each mean. As CME width 
increases, so does the mass of the CME (see Section \ref{S-widthflareflux}). 
}
\label{F-cwidthmass}
\end{figure}

Our method for pairing flares and CMEs creates a very large data set with which to statistically 
examine whether flare and CME parameters are correlated. With this data set, we have found 
correlated properties of associated flares and CMEs which are consistent with previous observations, 
and we have furthermore shown a relationship between the flux of the flare and the mass of the CME. 
In summary, we have shown:
\begin{enumerate}
\item{} CMEs associated with flares have higher average linear speeds than CMEs not associated with 
flares (Section \ref{S-flarecmespeed}; 495$\pm$8 km s$^{-1}$ and 422$\pm$3 km s$^{-1}$, respectively),
\item{} CMEs with higher linear speeds tend to decelerate; the relationship of acceleration and linear 
speed is correlated, and monotonically decreasing (Section \ref{S-flarecmespeed}),
\item{} Flare-associated CMEs tend to decelerate: flare associated CMEs have a mean acceleration of 
-1.8$\pm$0.1 m s$^{-2}$ while CMEs not associated with flares have a mean acceleration of 
0.07$\pm$0.25 m s$^{-2}$ (Section \ref{S-cmassaccelfflux}),
\item{} The magnitude of a flare-associated CME's deceleration increases with increasing flare flux: 
M class and stronger flare-associated CMEs have a mean acceleration 
of -3.7$\pm$1.3 m s$^{-2}$, while B class and weaker flare-associated CMEs have a mean acceleration of 
0.14$\pm$0.8 m s$^{-2}$ (Section \ref{S-cmassaccelfflux}),
\item{} A CME's mass is apparently unrelated to its acceleration (Section \ref{S-cmassaccelfflux}),
\item{} The width of a flare-associated CME is directly correlated with the flux of the flare, with 
X class flare-associated CMEs being the widest (80\degr$\pm$10\degr) and B class flare-associated 
CMEs the most narrow (42\degr$\pm$1.4\degr),
\item{} Wider CMEs are more massive than narrow CMEs (Section \ref{S-widthflareflux}, Figure 
\ref{F-cwidthmass}), and
\item{} The stronger the associated flare is, the more massive the CME is (Section \ref{S-massflux}); 
quantitatively, we find log(CME mass) $\propto$ 0.7$\times$ log(flare flux).
\end{enumerate}

\begin{figure}
\centerline{\includegraphics[width=5.5in]{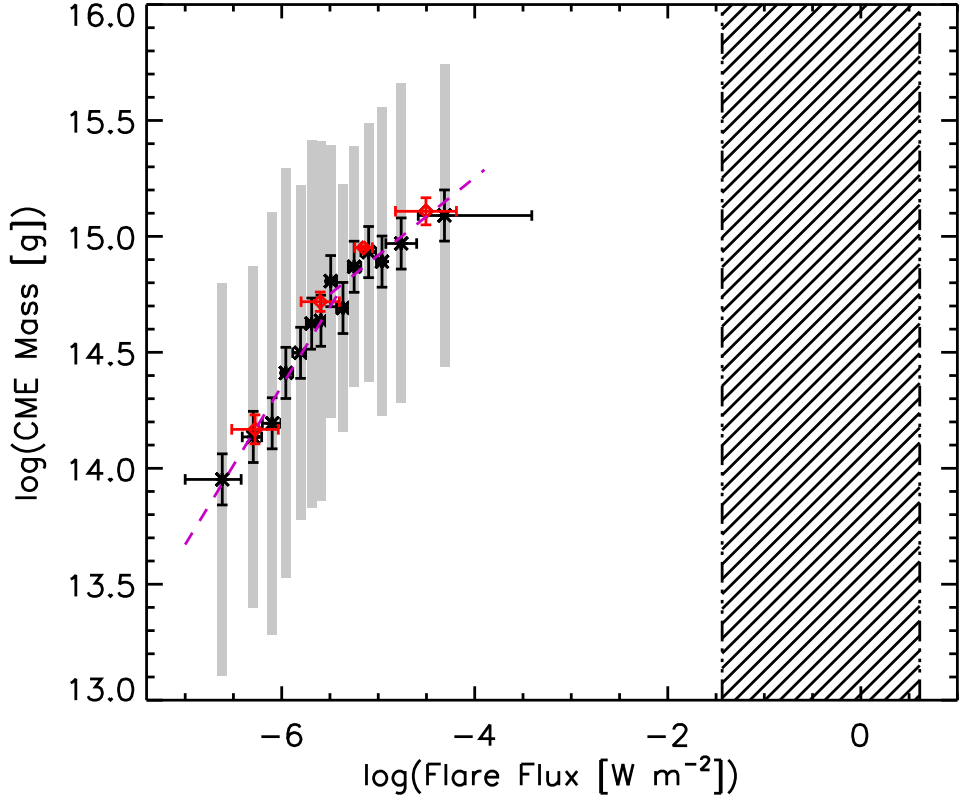}}
\caption{Relationship between CME mass and flare flux. 
Black exes:  the 826 flare/CME pairs are binned into equivalent boxes of $N =$~59. Each 
asterisk point is centered on the mean flux and mass per 59 pair bin.  The flux error bars 
shows the minimum/maximum flare flux values spanned by that bin, and the mass error bars are 
the error on the mean (\textit{i.e.}, $\sigma$/$\sqrt{N}$).
Four red diamonds:  abscissae are the mean flux values 
for each of the groups as set in Figure~\ref{F-gaussfits}; $\pm$x error is the standard deviation 
of that mean.  The ordinates are the mass values corresponding to the peaks of the Gaussian 
fits in Figure~\ref{F-gaussfits}; their errors are the fit errors of the centroid. 
The light gray shaded boxes in the background are of arbitrary width, but show 
the standard deviation of the mean mass plotted in the foreground.
For comparison, the hatched region represents the highest flux flares observed on young stars 
(see Section \ref{intro}).
Two linear functions are fit to the data, the first to bins 0-8, and the second to bins 7-13. 
There is slight overlap in that precisely where the ``knee'' of the function occurs is unclear. 
The first linear fit is of the form 
log(CME mass)=(18.5$\pm$0.57)$+$(0.68$\pm$0.10)$\times$log(flare flux). 
The second linear fit follows the form 
log(CME mass)=(16.6$\pm$1.30)$+$(0.33$\pm$0.26)$\times$log(flare flux).
}
\label{F-massfluxrel}
\end{figure}

\begin{figure}
\centerline{\includegraphics[width=5.5in]{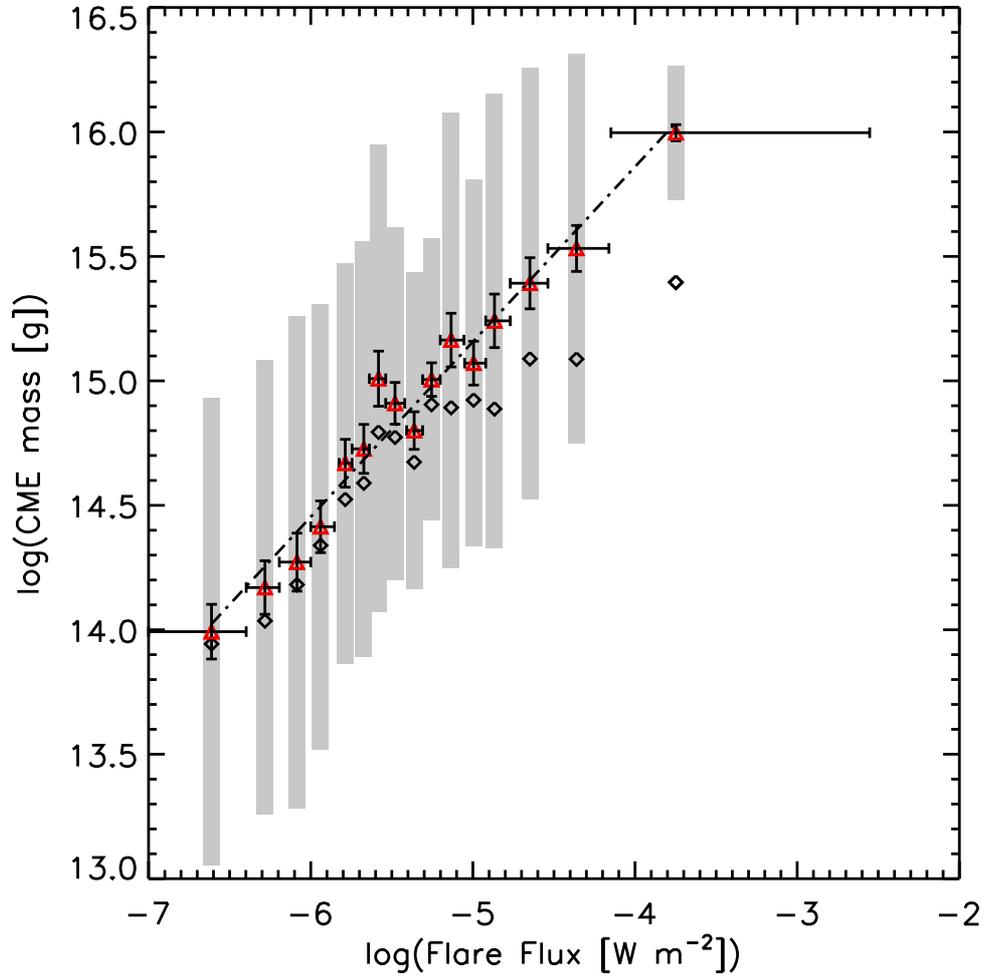}}
\caption{As in Figure \ref{F-massfluxrel}, the relationship between flare flux and CME mass 
is shown. In this case, we include the sample of CMEs with poorly constrained mass 
measurements (Figure \ref{F-masshist}, dashed histogram), and see a relationship (diamonds, black) 
much like the one in Figure \ref{F-massfluxrel}. The triangles (red) indicate how the function 
would change were we to include halo CMEs; we assign halo CMEs the maximum mass from 
the bin occupied by its associated flare. This function, fit by the dot-dashed (black) line, 
represents the steepest slope we could anticipate assuming halo CMEs have masses which lie 
within the observed distribution.
} 
\label{F-massfluxrelUB}
\end{figure}

The final bullet point in the list above could have very interesting implications for studies of 
young, solar-type stars, which exhibit high-frequency, high energy X-ray flaring activity. With the 
functional relationships illustrated in Figures \ref{F-massfluxrel} and \ref{F-massfluxrelUB}, it 
is now possible to extrapolate a CME mass/flare flux function into the range of flare fluxes observed 
for young stars (hatched region, Figure \ref{F-massfluxrel}). Whether the second function of the 
broken log-linear relationship (Figure \ref{F-massfluxrel}) or the single function of the analysis 
including halo and poorly measured CME masses (Figure \ref{F-massfluxrelUB}) would be used in this 
extrapolation remains to be seen; additional analyses of halo CME masses will be necessary to determine
whether the ``knee'' exists for physical reasons and remains present when halo CMEs are included with 
better mass estimates. If the change in slope is physical, it could potentially represent a saturation 
point, approaching a set limit of possible CME mass. For young stars, it would be interesting to see 
whether this saturation point has any correlation with the saturation point in X-ray activity.

\begin{acks}
This research is supported by NSF grant AST-0808072 (K. Stassun,
PI). K.~G.~S. gratefully acknowledges a Cottrell Scholar award from the
Research Corporation, and a Diversity Sabbatical Fellowship from the
Ford Foundation. Work at Boston University was supported by CISM which 
is funded by the STC Program of the National Science Foundation under 
Agreement Number ATM-0120950.
\end{acks}

\clearpage

\bibliographystyle{spr-mp-sola-cnd}
\bibliography{refs_cfc2} 

\end{article}

\end{document}